\documentclass{iau307}
\usepackage{graphicx}
\usepackage{natbib}
\usepackage{url}
\bibpunct{(}{)}{;}{a}{}{,}

\vspace{1cm}

\title[Stellar rotation] 
{Physics of rotation: problems and challenges}

\author[Maeder and Meynet]   
{Andre Maeder
 \and Georges Meynet}

\affiliation{Geneva Observatory, University of Geneva \\ email: {\tt andre.maeder@unige.ch} \\[\affilskip]
}

\pubyear{2014}
\volume{307} 
\pagerange{}
\jname{New windows on massive stars: asteroseismology, interferometry, and spectropolarimetry}
\editors{G. Meynet, C. Georgy, J.H. Groh \& Ph. Stee, eds.}

\begin{document}

\maketitle

\begin{abstract}
We examine some debated points in current discussions about rotating stars: the shape, the gravity darkening, the
critical velocities, the mass loss rates, the  hydrodynamical instabilities, the internal mixing and N--enrichments. The  study of rotational mixing requires
high quality data and careful analysis. From recent studies where such   conditions are fulfilled,   rotational  mixing is well confirmed.  Magnetic coupling with stellar winds may produce an apparent contradiction, i.e. stars with a low rotation and a high N--enrichment. We point out that it rather confirms  the large role of shears in differentially  rotating stars for the transport processes. New models of interacting binaries  also show how
shears and mixing may be enhanced   in close binaries  which are either spun up or down by tidal interactions.

\end{abstract}

\firstsection 

\section{Introduction}

Interferometry, asteroseismology and spectropolarimetry have brought  new  evidences  about the high impact of stellar rotation
on the stellar structure and evolution. We can say that rotation influences all observational properties as it has an impact on all
model outputs, whether for single or binary stars. The question is whether models and observations are in
agreement.
We concentrate in this review on current problems and new challenging questions regarding the effects of axial rotation
on stellar structure and evolution. For basic developments on the effects of rotation on stellar structure, evolution and nucleosynthesis, the reader may see for example \citet{MM2012}.

\section{Shape, gravity darkening, critical velocities
and mass loss \label{SecOne}}

\subsection{Shape of rotating stars}
The classical Roche model assumes that the gravitational potential is only due to a central mass concentration. At critical rotation,
i.e. when the outwards centrifugal force just compensates central gravity, the Roche model leads to an extreme ratio of the equatorial radius to the polar radius
 $R_{\mathrm{e}}/R_{\mathrm{p}} =1.5$.  (Structure models predict that  the polar radius generally decreases
by a few percents for extreme rotation, but this does not affect the critical ratio
 $R_{\mathrm{e}}/R_{\mathrm{p}} $). 

The VLTI observations of fast rotating stars have led to many discussions, particularly in the case of the Be star Achernar.  \citet{Domiciano2003} first
found a value of  1.56 for the ratio of the equatorial to the polar radius of Achernar, which was a problem for the Roche model. \citet{Kervella2006}
have studied the oblateness of Achernar and shown that the observations are influenced by the presence of a circumstellar envelope along the polar axis, in addition to the rotational flattening of the photosphere. \citet{Carciofi2008} pointed out that the controversial observations may be better
interpreted  with the account of gravity darkening with in addition a small equatorial disk making the transition between the photosphere and the circumstellar environment. \citet{Delaa2013}
also demonstrated in the case of $\alpha$ Cep  the importance of a good determination of the position angle of the rotation axis, in addition to the other mentioned effects.

On the theoretical side, \citet{Zahn2010} went a step beyond the Roche model by accounting  for the quadrupolar moment of the mass distribution for a star of 7 M$_{\odot}$ corresponding to Achernar. They showed 
that at critical velocity, the ratio  $R_{\mathrm{e}}/R_{\mathrm{p}} $ exceeds the standard value of 1.50. In the case of uniform rotation, they found that the extreme ratios $R_{\mathrm{e}}/R_{\mathrm{p}} $ ranges from 1.526 to 1.516
as the star evolves on the Main Sequence (MS). In the case of shellular rotation (with angular velocity $\Omega$
constant on level surface and increasing with depth), the values range from 1.560 to 1.535 over the MS phase. Thus, accurate interferometric observations
of stars at critical rotation might potentially  provide internal constraints on their internal rotation.

\subsection{Gravity darkening}
The von Zeipel theorem  \citep{vonZeipel1924} states that the flux  $\vec{F}(\Omega,\vartheta)$  at given angular velocity $\Omega$
and colatitude $\vartheta$ 
on a uniformly rotating varies like the effective gravity $\vec{g}_{\mathrm{eff}} (\Omega,\vartheta)$ , which is the sum of 
the Newtonian gravity and of the centrifugal acceleration. The von Zeipel theorem in the case of shellular rotation leads to \citep{Maeder1999}, 
 \begin{eqnarray}
 \vec{F}(\Omega,\vartheta)  =  - \frac{L} {4 \, \pi \, G \, M^{*}} \; \vec{g}_{\mathrm{eff}} (\Omega,\vartheta) [1+\zeta(\Omega,\vartheta)] \quad
 \label{vonzeipel}  
 \mathrm{with} \quad 
  M^{*} =  M \left(1 - \frac{\Omega^2}{2 \, \pi \, G  \, \overline{\varrho}_M} \right) .
  \label{mstar}
  \end{eqnarray}
\noindent
$L$ is the luminosity and $ \overline{\varrho}_M$ the average density over the mass $M$. The reduced mass $M^{*}$ was generally forgotten in
previous studies, despite the fact that it should also be there in case of uniform rotation. The  term $\zeta(\Omega,\vartheta)$ is only present in differentially rotating stars, it  brings correcting terms depending on the $\Omega$--gradient, 
the opacities and the gradient of $\mu$ (which also depends on ionization). Without the term  $\zeta$ and the mass reduction, Eq. (\ref{vonzeipel}) implies that $T_{\mathrm{eff}}$ behaves likes $g_{\mathrm{eff}}^{\beta}$, with $\beta=0.25$ in the classical case. \citet{Claret2012} has also found significant deviations from  the classical case, which depend on the optical depth, on $T_{\mathrm{eff}}$ and on the adopted atmosphere model.

Several authors have attempted to determine the parameter $\beta$ from interferometric observations, for recent references see \citet{Zhao2009,Che2011,Delaa2013}. Che et al. support a value $\beta=0.19$ for stars with $T_{\mathrm{eff}} > 7500$ K. Below, convective
envelopes tend to appear and imply low $\beta$-value as shown long ago by \citet{Lucy1967}.

We emphasize that gravity darkening affects all photometric and spectroscopic  observations. A rotating star may be seen as a composite
star made of thousands local atmosphere models with different local values of $g_{\mathrm{eff}}$ and $T_{\mathrm{eff}}$. Gravity darkening says how these parameters are distributed over the stellar surface. In addition, all these local models are seen with different limb--darkening effects. 

\begin{figure}[t]
\begin{center}
\includegraphics[width=9.5cm]{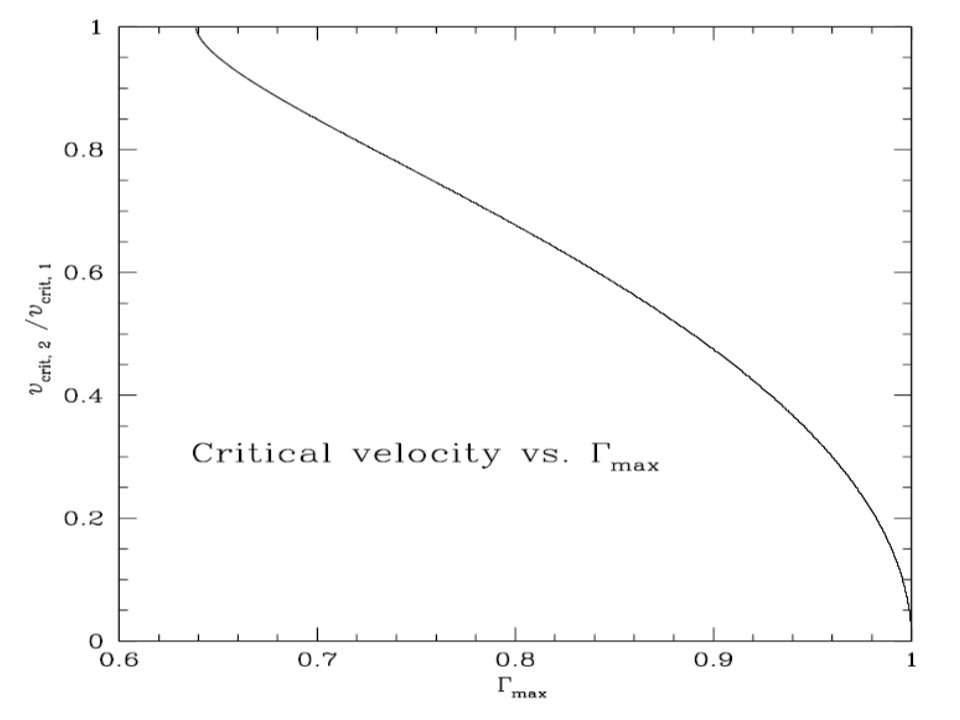} 
\caption{The ratio  $(v_{\mathrm{crit, 2}}/v_{\mathrm{crit, 1}})$ of the second root of Eq. (\ref{gg0}) to the first root as a function of $\Gamma_{\mathrm{max}}$ (the maximum  over the surface).  Eq. (\ref{v2v1})  describes this relation to better than 1\%.}
\label{vcrit2}
\end{center}
\end{figure}

\subsection{Critical velocities}

The classical expression of the critical or break--up velocity of a rotating star is 
\begin{eqnarray}
v_\mathrm{crit, 1} \,  = \,\left(  \frac{GM}{R_\mathrm{e,crit}} \right)^{\frac{1}{2}} \, = \,
\left( \frac{2}{3} \frac{GM}{R_\mathrm{p,crit}} \right)^{\frac{1}{2}} \; .
\label{vcrit1}
\end{eqnarray}
In massive stars, the high radiation pressure may add its outwards force to the centrifugal force and modify the expression of the critical
velocity. One often finds in literature the following expression with the Eddington factor $\Gamma$,
\begin{eqnarray}
v_{\rm{crit}} \, = \, \sqrt{\left(\frac{ GM}{R_{\mathrm{e, crit}}} \right)  (1-\Gamma)}  \; ,
\label{glanger}
\end{eqnarray}
\noindent
which is erroneously supposed to account for the reduction of the critical velocity by radiation pressure. Indeed, this expression 
ignores the von Zeipel theorem, which decreases the effect of radiation pressure at equator.

To obtain a correct expression,  one has first to  express the Eddington factor in a rotating star, accounting that both the actual radiation flux and the limiting maximum flux are influenced by the von Zeipel effect. This leads to  \citep{maederlivre09}
\begin{eqnarray}
\Gamma(\Omega,\vartheta) =
\frac{ \kappa (\Omega,\vartheta) \; L}{4 \pi \,
c\, G\,M \left( 1 - \frac{\Omega^2}{2 \pi G \overline{\varrho}_{M}}  \right) } \; .
\label{gammaomega}
\end{eqnarray}
\noindent
This is the local Eddington factor, which varies as a function of rotation and colatitude $\vartheta$. Now, the correct expression of the critical
velocity is  defined by
\begin{eqnarray}
 \vec{g_\mathrm{eff}} \;\left[1 - \Gamma_{\Omega}(\vartheta)
 \right] = \vec{0} .
 \label{gg0}
\end{eqnarray}
This is a second degree equation, which thus has 2 roots. For each couple $(\Omega,\Gamma)$, one has to consider the smallest of the two roots.
For $\Gamma < 0.639$, the smallest root is just given by Eq. (\ref{vcrit1}). Due to the von Zeipel theorem, the radiative flux at equator decreases
in the same way as the effective gravity as rotation increases. The second root has a relatively complex expression \citep{maederlivre09}, which is illustrated in Fig. \ref{vcrit2}.

We see that for an Eddington factor tending towards 1, the critical velocity tends to zero, because  even for a zero velocity 
the surface is unbound. This  reduction of the critical velocity  comes from the reduction of the effective mass in
Eq. (\ref{mstar}). The exact expression of $v_{\mathrm{crit, 2}}$, given by Eq. (4.42) in \citep{maederlivre09}, is a bit difficult to handle.
A very good fit of the line in Fig. \ref{vcrit2} can be made by a combination of a linear  and an exponential relation, 
\begin{eqnarray}
\frac{v_{\mathrm{crit,2}}}{v_{\mathrm{crit,1}}}=1.00 -1.93 \, (\Gamma-0.63)-0.286 \; \Gamma^{40}.
\label{v2v1}
\end{eqnarray}
For $\Gamma > 0.639$, it gives a fit with an accuracy better than 0.01.

\subsection{Mass loss}
How does rotation affect mass loss? We are now facing two opposite answers to this important question. From the application of the force multipliers, an increase of the mass loss rates  $\dot{M}$ with $\Omega$  is predicted \citep{maederlivre09}, 
\begin{eqnarray}
\frac{\dot{M}(\Omega)}{\dot{M}(0)}= \left[1 - \frac{4}{9} \left( \frac{v}{v_{\mathrm{crit,1}}} \right)^2 \right]^
{\frac{7}{8}-\frac{1}{\alpha}} \; .
\label{mdot}
\end{eqnarray}
\noindent
The force multipliers  take values $\alpha$=0.64, 0.64, 0.59, 0.565 for $T_{\mathrm{eff}}$=50'000, 40'000, 30'000, 20'000 K
\citep{Pauldrach1986}. The  term $\frac{4}{9} \left( \frac{v}{v_{\mathrm{crit,1}}} \right)^2$ comes from $\frac{\Omega^2}{2 \pi G \overline{\varrho}_{M}}$ in $M^{*}$.

Oppositely, a decrease of the mass loss  with rotation is predicted by \citet{Muller2014}, who   apply a parameterized description of the line acceleration that only depend on radius at a given latitude.  We note that they use the incorrect expression  (\ref{glanger}) of the
critical velocity and do not account for the mass reduction $M^{*}$. Whether, this explains the difference in their conclusions is uncertain.
In this context, we note that there are enough data on rotational velocities and mass loss rates to allow an observational test.

\section{A new global approach about mixing}

\begin{figure}[h]
\begin{center}
\includegraphics[width=9cm,height=6cm]{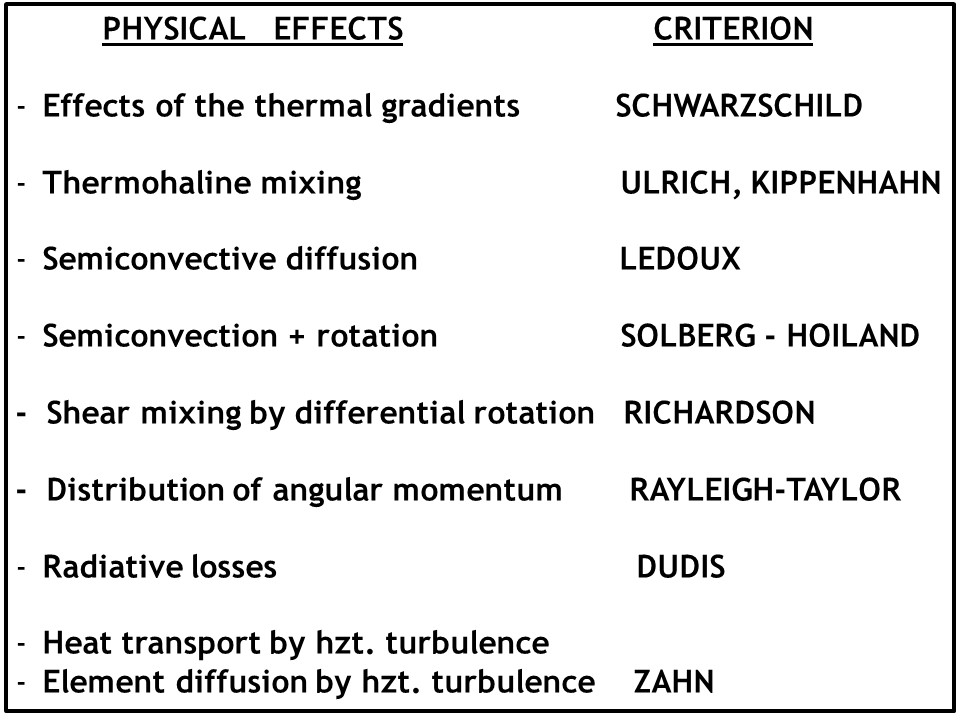} 
\caption{The hydrodynamical effects accounted for in the new criterion.}
\label{effects}
\end{center}
\end{figure}

The instabilities producing chemical mixing and transport of angular momentum, such as those listed in Fig. \ref{effects},  are generally considered separately and the global diffusion
 diffusion coefficient is taken as the sum of the various particular coefficients. This is incorrect and ignores the interactions between the various instabilities, which can amplify or damp each other.

A new stability condition may be written, accounting for the various instabilities mentioned \citep{Maeder2013}. It  is a quadratic equation of the form $Ax^2+Bx+C > 0$,
\begin{eqnarray}
\left[N^2_{\mathrm{ad}}+N^2_{\mu}+N^2_{{\Omega}-\delta v}
\right]\, x^2+ \nonumber \\
\left[N^2_{\mathrm{ad}}D_{\mathrm{h}}+N^2_{\mu}(K+D_{\mathrm{h}})+ 
N^2_{{\Omega} -\delta v}
(K+2 D_{\mathrm{h}}) \right] \, x+ \nonumber \\
N^2_{{\Omega}-\delta v}
(D_{\mathrm{h}}K+
D^2_{\mathrm{h}}) \, > \,0 \; , 
\label{2ndeq}
\end{eqnarray}
\noindent
\begin{equation}  
\, \mathrm{with} \quad  \; N^2_{{\Omega}-\delta v} = \frac{1}{\varpi^3} \, \frac{d\left( \Omega^2 \, \varpi^4 \right) }{d\varpi} \, \sin \vartheta
 - \mathcal{R}i_{\mathrm{c}} \left(\frac{dv}{dr}\right)^2 .
\end{equation}
\noindent
Expression of $N^2_{{\Omega}-\delta v}$
 is a modified form of the Rayleigh oscillation frequency, accounting for both  the angular momentum  distribution and the excess of energy in the shear.  The quantity $\varpi$ is the distance to the rotation axis, ${K}$ the thermal diffusivity and $D_{\mathrm{h}}$ the diffusion coefficient of 
horizontal turbulence. Various forms of  $D_{\mathrm{h}}$  are existing \citep{maederlivre09}.
Eqn. (\ref{2ndeq}) is the general equation, which should 
 be considered in a rotating star to account for the various effects mentioned above. {\emph{The global diffusion coefficient $D_{\mathrm{tot}}$
is just given by  $D_{\mathrm{tot}} \, = \, 2  x $ and not by the sum of the specific coefficients}}. The solution of an equation of the second  degree
is not the sum of some peculiar solutions !

We notice several interesting consequences. - All usual criteria need to account for rotation. - The horizontal turbulence has an effect on
various transport processes. - Interestingly enough, a stable distribution of angular momentum according to the Rayleigh--Taylor
may reduce the effect of a shear. - Conversely, the simultaneous account of shears may also enhance the Rayleigh--Taylor instability.
Globally, the interactions of the various effects may either damp or enhance the instabilities.

\section{Rotation and  N-enhancements: on some confusions}

\begin{figure}[h]
\begin{center}
\includegraphics[width=13cm,height=7cm]{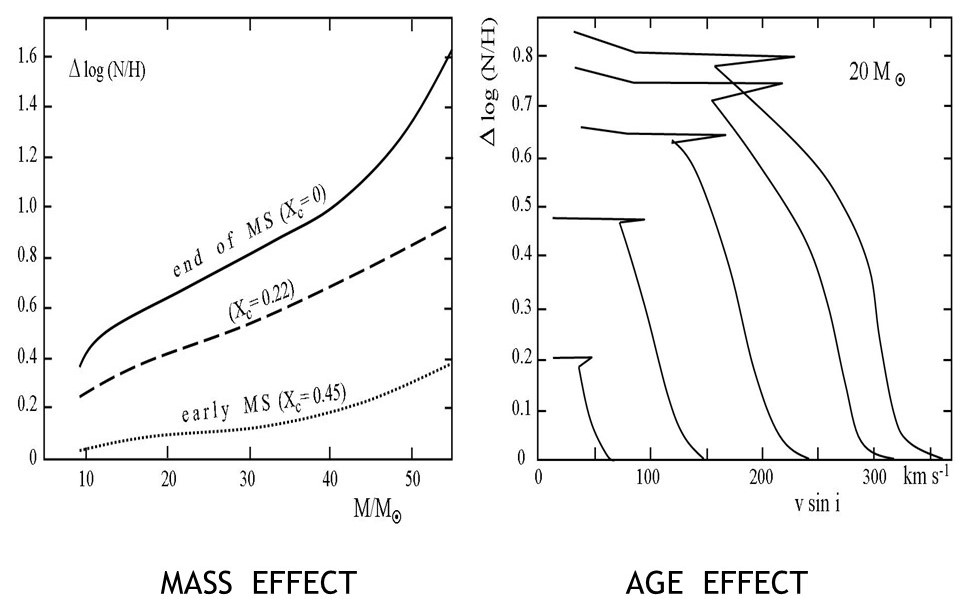} 
\caption{Left: The relative variations of  log $(N/H)$ for average rotational velocities as a function of the initial stellar masses of model at $Z=0.02$
at 3 different evolutionary stages on the MS labeled by $X_{\mathrm{c}}$ the central H--content. Right: The relative variations of log $(N/H)$ as function of the rotational
velocities along the MS phase, starting from different initial velocities, in the case of a model of 20 M$_{\odot}$ with Z=0.02 \citep{Maeder2009}.}
\label{mass}
\end{center}
\end{figure}

\subsection{The physical effects influencing chemical mixing}

The relations between rotational  velocities  and N--enrichments have recently been the object of many confusing statements. It is thus necessary to clarify the matter. Let us first say a few words about the physical effects which enter into the game. 
The models of rotating intermediate and massive stars  made since 15 years clearly show the role of several effects, confirmed by more recent works   \citep{Grille2012,Georgy2013}, as well as by observations \citep{Bouret2013}. In particular, {\emph{the N-enrichments resulting from internal mixing
are a multivariate function of several parameters: first of the  rotation velocity, but  also of the mass, age, binarity, metallicity and  magnetic fields}}. Let us note that some models of rotating stars consider  the transport (particularly of angular momentum) by meridional circulation as a diffusion, while it is an advection. The two processes are physically not the same: a diffusion transports something from a region where there is a lot to a place where there is very little. An advection (or circulation) can transport matter in any directions,
even from where there is a little to a place where there is a lot.
Thus, treating meridional circulation as a diffusion, as often done,  can even give the wrong sign for  the transport !

Besides the rotational velocities which directly influence mixing, there are several other parameters which play a role. -- 1. Mass: 
Fig. \ref{mass} (left)  shows the variations of N--enrichment with stellar mass for an average rotation velocity. Depending on the masses considered, the enrichments may vary by more than an order of magnitude. 
--2. The age effect is illustrated on the right: we see that for a given mass and for the same observed velocity, the enrichments
may also differ by more than an order of magnitude depending on the
evolutionary stage considered \citep{Maeder2009}. 

\begin{figure}[h]
\begin{center}
\includegraphics[width=9cm]{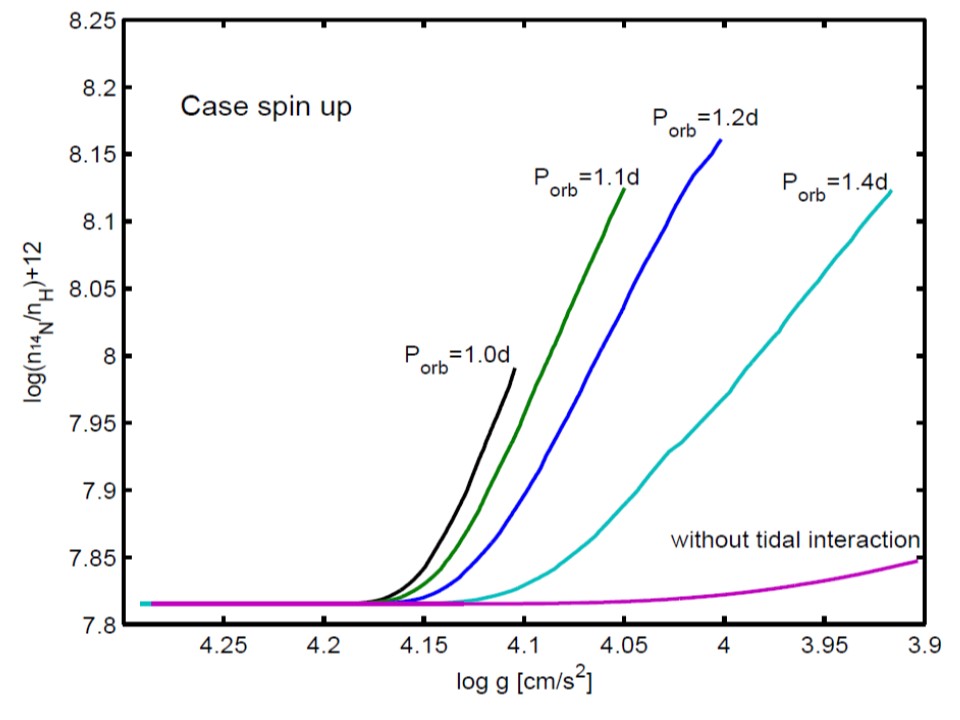} 
\caption{Variation of the N content as a function of log $g$ over the MS phase for the primary of a binary system of  a 15  + 10 M$_{\odot}$ 
with different initial orbital periods $P_{\mathrm{orb}}$ which spin--up during the MS phase starting from an initial rotation velocity equal to 20\% of the critical value. The mixing is mainly due to tidally induced shear mixing \citep{Song2013}.  }
\label{Song}
\end{center}
\end{figure}

--3. The tidal interaction in binaries, even before mass transfer, may also drastically 
enhance the N--enrichments \citep{Song2013}.  Fig. \ref{Song} illustrates a case of tidal interaction in binaries, where the primary is spun--up from a low initial rotation velocity by tidal interaction. The N--enrichments are considerable.  The tides accelerate the external layers, this  increases the shear  throughout the star and thus the N--enrichments at the surface. The opposite case may also occur  \citep{Song2013}: the tides may produce a spin--down of initially fast rotators, the resulting shears also contribute to N--enrichments. Works 
in progress show that due to tidal mixing some stars may even become chemically homogeneous, thus they evolve contracting and  never reach
a stage of mass transfer.  

--4. Metallicity $Z$  influences  mixing:  low $Z$ stars being more compact experience stronger shears, which favor mixing and
 N--enrichments \citep{Maeder2001}. The effect applies down to very low metallicities, as shown by current
studies on the origin of the primary nitrogen. In addition, there are clear indications that at lower metallicities, stars have on the average  faster 
rotational velocities  than at solar composition \citep{MaederGM99,Martayan2007}. This makes the rotational mixing even a more important
effect in lower metallicity populations and earlier galactic evolutionary stages.

 --5. The relation between magnetic field and N--enrichments is complex 
both theoretically and observationally. Models with the Tayler--Spruit dynamo predict significant N--enrichments, however  
 the existence of this dynamo in radiative zones of differentially rotating stars is still very debated \citep{Zahn2007}.  The  observations indicate that some magnetic stars are N--enriched and some are not
(cf. presentations by Neiner,  by  Henrichs,  by Wade at this meeting). It is clear that magnetic fields transport the angular momentum and favor the internal coupling of the rotation, but as far as the chemical mixing is concerned  the situation remains uncertain.
Things are more clear concerning the effects of magnetic fields at the stellar surface, which
are  observed  in about 10\% of the  massive stars. These fields produce  a mechanical  coupling of the stellar surface with the stellar winds. This  removes angular momentum from the star \citep{Uddoula2002,Uddoula2008}. The prescriptions of these authors were used in models  by 
 \citet{Meynet2011}, who showed that the magnetic coupling can produce an efficient braking.  This braking at the surface   considerably  increases the internal shear 
and thus makes a strong mixing. This leads to massive stars with low rotation velocities and at the same time strong enrichments.
Thus, the surface effects of magnetic fields may produce stars which just seem to contradict the current expectations about a simple relation between N-enrichments and $v \sin i$. However, it rather confirms  the large role of shears in differentially  rotating stars for the transport processes. 

As a conclusion, for a careful observational study of the rotational mixing, it is
indispensable to strongly limit the range of the other intervening parameters, mass, age, binarity, metallicity and  magnetic filed.

\subsection{Some current observations on N-enrichments}

The VLT--Flames survey of massive stars in the Magellanic Clouds  lead  the authors to challenge the existence of rotational mixing
\citep{Hunteretal07,Hunteretal09,Brottetal2011}. They found two groups of stars, which do not follow a relation
between $v \sin i$ and $N/H$ ratios. Group 1 (25 \% of the stars)
with little enrichments and high $v \sin i$ and Group 2 (16 \% ) with
high enrichments  and low $v \sin i$.   The sample was rediscussed by  \citet{Maeder2009}, who showed that the sample  by Hunter et al. is in fact a mixture of masses between 10 and 30 $M_{\odot}$
and of evolutionary stages, with stars close to the Main--Sequence and stars out of it. When a  limited range of masses and
ages  is analyzed, the correlation improves considerably \citep{Maeder2009}, as far as the accuracy of the data permit it. Group 1 essentially vanishes, while Group 2 may be
explained by the effect of magnetic braking mentioned above, it also contains some  evolved stars.

Regarding possible tests about the accuracy of the chemical abundances, the properties of the N/C vs. N/O diagram
have recently been discussed by \citet{MaederCNO2014}. They show that up to an enrichment of N/O by a factor of 4, the  relation between N/C and N/O is essentially model independent, being determined only by nuclear physics.  Up to the mentioned enrichment, the scatter around the mean relation,
if anyone, is essentially observational or coming from the analysis process. Such a test beautifully confirms the high quality of the data
 by \citet{Przy2011}, while the quality of the abundance data from \citet{Hunteretal07,Hunteretal09} appears much lower.

A detailed analysis  of the chemical abundances of O--stars in the SMC was made by \citet{Bouret2013}.  They found that 65\% of the observed stars well fit the predicted relations of N--enhancement versus $v \sin i$, 26\% are marginal and  9\%  do not fit. This fraction of outliers is much lower than that found by  \citet{Hunteretal09} and \citet{Brott2011}. In particular, Bouret et al.  confirm the existence of  Group 2 with slowly rotating N--rich stars. As we have stressed, this group may result fom magnetic braking. They confirm that faster rotators have higher N/C ratios, that more massive stars have higher N/C than less massive stars. The same is true for the N/O ratios.

Finally, we must also mention a recent study of 68 bright stars of all over the sky
with high--precision spectrocopic and asteroseismic measurements by \citet{Aerts2014}. These authors conclude "We deduce that neither the projected  rotational velocity nor the rotational frequency has predictive power for the measured nitrogen abundance". An examination 
of the data used in this analysis shows stars with spectral types from A0 to O5, i.e. with masses between about 3 and 30 M$_{\odot}$.  Fig. 3 left
indicates that this range of masses  is much too large to allow reasonable comparisons.
 In addition, the sample contain stars of all luminosity classes: dwarfs, giants and supergiants, which should also lead to some difficulties in the comparisons according to Fig. 3 right. These  stars are all over the sky, spanning a range of galactocentric distances of several kpc, thus with significant metallicity differences. Therefore,  the conclusions by Aerts et al. regarding the the effects of rotation on nitrogen 
abundances are a little premature.

\section{Conclusions: the next great challenge}

We hope that the above lines help to clarify some points in current discussions about rotating stars, concerning their shape, gravity darkening, 
critical velocities, mass loss, instabilities, mixing and N--enrichments. These points must not prevent us to see another big problem we
are now facing concerning rotation.

Recent asteroseismic observations   indicate  too slow rotation of stellar cores in red giants compared to model predictions. The precise 
determinations from KEPLER of the rotational frequency splittings of mixed 
modes provide information on the internal rotation of red giants. \citet{Beck2012}
 showed that in the case of the red giant KIC 8366239 the core rotates about 10 times faster than the surface. 
\citet{Deheuvels2012,Deheuvels2014} analyzed the rotation profiles of six red giants and  found that the cores  rotate between   3 and 20  times faster than the envelope. They also found that the rotation contrast between  core and envelope increases during the subgiant branch. 
The crucial point is that the  observed differences of rotation between cores and envelopes are much smaller than predicted by the evolutionary models of rotating stars. Analyses with stellar modeling by several  groups  \citep{Eggenberger2012,Marques2013,Ceillier2013,Tayar2013} clearly demonstrate that some
 generally unaccounted  physical process is at work producing an important internal viscosity. 

In this context, we must also remember previous results showing that
the evolutionary models leading to pulsars always  rotate much too fast
compared to the observed rotation rates of pulsars \citep{Heger2004,Hirschi05}. This is even the case when the magnetic field of the Tayler--Spruit dynamo \citep{Spruit2002}, which could act
in rotating radiative regions, is accounted for.  
 A similar problem appears for white dwarfs, which generally show rotation velocities  much lower than predicted by standard models  \citep{Berger05}. Thus, there 
is an ensemble of observations pointing in favor of some additional internal coupling in star evolution.

These new windows provided by recent technologies are a  gift for our better understanding of stellar physics. To be fully enjoyed, the gift
must receive great care in reduction and analysis.

\bibliographystyle{iau307}
\bibliography{MyBiblio}

\begin{discussion}

\discuss{Wade}{In our poster we describe precise magnetic field measurements of the sample of 20 early B stars for which Nieva and Przybilla obtained high-quality nitrogen abundances. Of the 5 stars we identify as N-rich, 3 are magnetic. But 2 are not, with upper limits below $10\,\mathrm{G}$. To make the situation more confusing, Thierry Morel has reported N abundances for 2 other early B/late O magnetic stars (NGC 2244-201, HD 57682) which seem to be normal. So there seem to be a great diversity of objects!}

\discuss{Maeder}{Your point is an important one. Some years ago, the N/H excesses were  considered as a signature of magnetic fields. Such results would have been in agreement with our evolution models based on the Tayler-Spruit dynamo. However, evidences seem to accumulate in favor of a fossil origin of magnetic fields (cf. Mathis). In this context, an explanation is more difficult since the magnetic field lines are frozen in the medium (if the field is large) and this does not seem to favor the transport.}

\discuss{Baade}{To supplement Gregg Wade's comment: quite some years ago, Johann Kolb and I tried to measure N abundances of broad-line Be stars \textit{relative to Bn stars}, which seem to rotate about as rapidly as Be stars but not suffer outbursts leading to the formation of circumstellar disks. Moreover, Be stars are the only group of stars investigated by MIMES and found to be 100\% free of detectable magnetic fields. Yet, we found Be stars more nitrogen rich than Bn stars. Unfortunately, this project remained unfinished since Johann left astronomy. As well, Thomas Rivinius told me that he and Maria-Fernanda Nieva measured N in some narrow-lined Be stars. So far, the stars studied have solar N abundances. Therefore, one of the two preliminary results is in error or the N abundances depend on stellar latitude.}

\discuss{Aerts}{How would the coupling of a magnetic field impact on the internal rotation profile for a B star with a radiative envelope? (see our latest $\Omega(r)$ profile deduced from seismology of a B star with gravity waves.)}

\discuss{Maeder}{The models we have at present are based on the Tayler-Spruit dynamo. They indicate that for the fields created by this dynamo, the internal profiles of $\Omega(r)$ are very close to solid body rotation. We have not yet models for weak fossil fields, but below some critical field value, it is likely that some significant differential rotation may exist.}

\discuss{Arlt}{It should be noted that magnetic instabilities deliver two different diffusion coefficients for angular-momentum transport and mixing. The slow-down and nitrogen peculiarities may therefore develop on different time-scales.}

\discuss{Maeder}{The equations of transport of the angular momentum and of chemical transport are different and the transport coefficients are not necessarily the same. However, this depends on the magnetic instabilities and horizontal turbulence.}

\discuss{Noels}{If most massive stars undergo this ``needed'' strong braking already in the early stages of the main sequence, what are the consequences on the nitrogen enrichments?}

\discuss{Maeder}{The external braking by magnetic fields enhances  differential rotation and thus the shears,  in turn these drive mixing and surface N-enrichment. On the contrary,  internal magnetic coupling generally reduces  the shears and thus the mixing.  The question is what is the balance between the internal and external effects. There are not yet  relevant 
models.  Thus, at the present time, it is difficult to answer your most interesting question. }

\end{discussion}

\end{document}